\documentclass[10pt,aps,prd,twocolumn,floats,floatfix,showpacs,superscriptaddress,nofootinbib]{revtex4-1}
\usepackage[utf8]{inputenc}               		

\usepackage{graphicx,mathtools,amssymb,amsmath,amsthm,amsfonts,epsfig,epsf}
\usepackage[outdir=./]{epstopdf}
\usepackage[linktocpage]{hyperref}
\hypersetup{hidelinks}
\usepackage[usenames]{color}		
\usepackage{tensor}	
\usepackage{csquotes}
\usepackage{mathrsfs}

\newcommand{\note}[1]{\text{\tiny{#1}}}

\newcommand{\dd}{\mathrm{d}}
\newcommand*\DAlembert{\mathop{}\!\mathbin\Box}

\begin{document}

\title{Spontaneous scalarization in generalized scalar-tensor theory}

\author{Nikolas Andreou}
\affiliation{School of Mathematical Sciences, University of Nottingham,
University Park, Nottingham NG7 2RD, United Kingdom}

\author{Nicola Franchini}
\affiliation{School of Mathematical Sciences, University of Nottingham,
University Park, Nottingham NG7 2RD, United Kingdom}

\author{Giulia Ventagli}
\affiliation{School of Mathematical Sciences, University of Nottingham,
University Park, Nottingham NG7 2RD, United Kingdom}

\author{Thomas P.~Sotiriou}
\affiliation{School of Mathematical Sciences, University of Nottingham,
University Park, Nottingham NG7 2RD, United Kingdom}
\affiliation{School of Physics and Astronomy, University of Nottingham,
University Park, Nottingham NG7 2RD, United Kingdom}

\begin{abstract}
Spontaneous scalarization is a mechanism that endows relativistic stars and black holes with a nontrivial configuration only when their spacetime curvature exceeds some threshold.
The standard way to trigger spontaneous scalarization is via a tachyonic instability at the linear level, which is eventually quenched due to the effect of nonlinear terms. In this paper, we identify all of the terms in the Horndeski action that  contribute to the (effective) mass term in the linearized equations and, hence, can cause or contribute to the tachyonic instability that triggers scalarization.
\end{abstract}

\maketitle

\section{Introduction}
Gravitational waves astronomy now provides a way to directly observe the strong-gravity corner of general relativity (GR). The principal source of gravitational waves observable by current detectors is compact binaries. Known detections include  ten black hole binary mergers and one merger where at least one of the two objects was a neutron star~\cite{LIGOScientific:2018mvr}. Moreover, the number of these observations will increase in the future, making it possible to constrain properties of black holes and neutron stars to unprecedented levels. While in GR compact objects are fairly well understood, this does not necessarily apply to modified theories of gravity. For example,  there is less than a handful of examples for binary merger simulations for non-GR black holes~\cite{Hirschmann:2017psw,Okounkova:2017yby,Witek:2018dmd}, and one needs concrete predictions to optimize constraints for any given theory.

A key question for black holes and neutron stars in modified theories of gravity is whether they carry some characteristics which make them distinguishable from their GR counterparts. For black holes, such characteristics are generally known as \textit{hair}. Significant effort  has been put into proving no-hair theorems and finding possible evasions, as a way to identify theories that can exhibit interesting strong-field phenomenology. For example, well-known no-hair theorems have been proven for scalar field nonminimally coupled to gravity under certain assumptions~\cite{Hawking:1972qk,Bekenstein:1971hc,Sotiriou:2011dz,Hui:2012qt,Silva:2017uqg}. For shift-symmetric scalars ({\em i.e.}~scalars that are protected from acquiring a mass from quantum corrections) it turns out that there is a unique nonminimal coupling term between the scalar and curvature that can lead to scalar hair~\cite{Sotiriou:2013qea,Sotiriou:2014pfa}. If one relaxes the assumption of shift symmetry though, no-hair theorems currently cover a limited subclass of theories, know as scalar-tensor theories (see~\cite{Faraoni:2004pi} for a review).

Interestingly, this is precisely the class of theories on which attention has been focused on the neutron star front. This is largely due to a specific model within that class  introduced by Damour and Esposito-Far\`{e}se (DEF) in~\cite{Damour:1993hw}. The model exhibits a phenomenon dubbed spontaneous scalarization: a linear tachyonic instability around a neutron star configuration that is a solution of GR (induced by the nonminimal  coupling between a scalar field and the metric) can trigger the growth of the scalar field~\cite{Harada:1997mr}. The instability is eventually quenched at nonlinear level and the end point is a neutron star ``dressed'' with a scalar configuration. The interesting part is that with a mild tuning of the parameters, one can set the threshold of this phase transition at typical densities of neutron stars. Thus, one can avoid scalarization in the Solar System, thereby evading weak-field constraints, while still having significant effects for neutron stars. In fact,  current constraints from binary pulsars almost rule out the original model because the deviation from GR would have already been observed~\cite{Shao:2017gwu}. However, one can circumvent such constraints by adding a small mass to the scalar~\cite{Ramazanoglu:2016kul}. A massive scalar can also resolve the tension between spontaneous scalarization and cosmological evolution~\cite{Alby:2017dzl}, avoiding initial data tuning~\cite{Damour:1992kf,Damour:1993id,Anderson:2016aoi}.

As already mentioned, the DEF model (with or without a mass) is covered by a no-hair theorem~\cite{Sotiriou:2011dz}, and hence it does not lead to black hole scalarization (see however~\cite{Cardoso:2013opa,Cardoso:2013fwa,Herdeiro:2019yjy}). It has been shown recently, though, that a different type of coupling between a scalar and curvature can lead to spontaneous scalarization for both neutron stars and black holes~\cite{Silva:2017uqg,Doneva:2017bvd}. Similarly to the DEF model, scalarization is triggered by a tachyonic instability and, as the scalar field grows, nonlinear terms eventually take over and quench the instability, thereby determining the properties of the final configuration~\cite{Silva:2018qhn,Macedo:2019sem}. These new models of scalarization are receiving a lot of attention lately (see {\em e.g.}~\cite{Silva:2017uqg,Doneva:2017bvd,Antoniou:2017acq,Blazquez-Salcedo:2018jnn,%
Minamitsuji:2018xde,Silva:2018qhn,Macedo:2019sem,Doneva:2019vuh}). This is justified because spontaneous scalarization is currently the only known mechanism that could make appear in the strong curvature fields that remain dormant at small curvature. As such, it is the most promising effective description for elusive new physics that could first make its appearance in gravitational waves observations. Indeed, it has been argued that if a scalar that undergoes such a phase transition is coupled to matter appropriately, then it could change the properties of the standard model within compact stars~\cite{Coates:2016ktu,Franchini:2017zzx}. Moreover, the mechanism of spontaneous scalarization can straightforwardly be extended to other fields~\cite{Ramazanoglu:2017xbl,Ramazanoglu:2018hwk,Ramazanoglu:2018tig,Ramazanoglu:2019gbz,Annulli:2019fzq} or other couplings~\cite{Herdeiro:2018wub}.

The new models of scalarization have clearly demonstrated that the DEF model is not unique in this respect. This suggests that there might be more, yet to be discovered, theories that exhibit spontaneous scalarization. Here we address this question for a scalar field that belongs to the Horndeski class. Horndeski gravity is the most general action one can write with a metric and a scalar field that leads to second order equations of motion upon direct variation~\cite{Horndeski:1974wa,Deffayet:2009mn}. Initially, Horndeski gravity (or its subclasses) gained a lot of interest as models of dark energy~\cite{Chiba:1999ka,ArmendarizPicon:2000dh,Ferreira:1997hj,DeFelice:2010pv,%
DeFelice:2010nf,Nojiri:2005vv,Koivisto:2006ai,Tsujikawa:2006ph,Nojiri:2005jg,Deffayet:2010qz,Kase:2018iwp}. However, the measurement of the speed of gravitational waves provided by the multimessenger neutron star binary merger has put severe constraints on such models \cite{Lombriser:2015sxa,Lombriser:2016yzn,Ezquiaga:2017ekz,Baker:2017hug,Creminelli:2017sry,Sakstein:2017xjx}. It should be stressed that these constraints rely crucially on the assumption that the scalar is cosmologically dominant and drives the cosmic expansion. Horndeski gravity is still viable and relevant for the description of compact objects if one assumes the scalar field to be subdominant in the cosmological evolution of the Universe \cite{Franchini:2019npi}.

We restrict our attention to the onset of scalarization and in particular to the conditions for scalarization to be triggered by a tachyonic instability. As discussed already above, even though nonlinearities are essential for determining the fate of the instability and pinning down the end state~\cite{Silva:2018qhn,Macedo:2019sem}, the onset of the instability can be captured in the linear regime already. This implies that one can obtain necessary conditions for spontaneous scalarization simply by inspecting the linearized field equations and the contributions to the effective mass term for the perturbation of the scalar.

Hence, in this paper we  proceed as follows.
In Sec.~\ref{Sec1} we give a brief overview of Horndeski theory, and we identify the condition that  the free functions of the theory should satisfy to have GR solutions admissible in general.  Then, we linearize the scalar field equation and we identify two different effective mass contributions for a scalar perturbation.
In Sec.~\ref{Sec3} we identify the minimal actions which contribute to each separate contribution to the mass term at linearized level. We then demonstrate that the two seemingly distinct contributions come from actions that are related by field redefinitions. We analyze each term in the minimal actions and we show how it relates to known scalarization models.
Finally, Sec.~\ref{Sec6} contains a discussion of our results.

\section{Horndeski gravity}\label{Sec1}
\subsection{The theory}

Horndeski theory is the most general action involving a scalar field that leads to second order field equations upon variation~\cite{Horndeski:1974wa}. The theory was rediscovered independently in the context of Galileons~\cite{Deffayet:2009mn}. The action of the theory can be written as~\cite{Kobayashi:2011nu}
\begin{equation}
\label{Horndeski}
S=\frac{1}{2\kappa}\sum_{i=2}^{5}\int \dd^4x\,\sqrt{-g}\mathcal{L}_i+S_{\note{M}},
\end{equation}
where we have defined
\begin{align}
\label{L2}
\mathcal{L}_2 = & \, G_2(\phi,X),\\
\label{L3}
\mathcal{L}_3 = & \, -G_3(\phi,X)\DAlembert\phi,\\
\label{L4}
\mathcal{L}_4 = & \, G_4(\phi,X)R+G_{4X}[(\DAlembert\phi)^2-(\nabla_\mu\nabla_\nu\phi)^2],\\
\notag
\mathcal{L}_5 = & \, G_5(\phi,X)G_{\mu\nu}\nabla^\mu\nabla^\nu\phi \\
\label{L5}
- & \frac{G_{5X}}{6}\left[\left(\DAlembert\phi\right)^3 -3\DAlembert\phi(\nabla_\mu\nabla_\nu\phi)^2+2(\nabla_\mu\nabla_\nu\phi)^3\right],
\end{align}
and $X=-\nabla_\mu\phi\nabla^\mu\phi/2$, $(\nabla_\mu\nabla_\nu\phi)^2=\nabla_\mu\nabla_\nu\phi\nabla^\mu\nabla^\nu\phi$, $(\nabla_\mu\nabla_\nu\phi)^3= \nabla_\mu\nabla_\nu\phi\nabla^\nu\nabla^\lambda\phi\nabla_\lambda\nabla^\mu\phi$ and $G_{iX}=\partial G_{i}/\partial X$. We have also defined $\kappa=8\pi G/c^4$ and $S_\note{M}$ is the matter action. Matter is assumed to couple minimally to the metric only and this means we are working in the so-called Jordan frame.

Varying the action with respect to the metric $g^{\mu\nu}$ and the scalar field $\phi$ yields respectively
\begin{align}
\label{HornEq1}
&\sum_{i=2}^{5}\mathcal{G}^{i}_{\mu\nu} =\kappa T_{\mu\nu}, \\
\label{HornEq2}
&\sum_{i=2}^{5}\left( P^i_\phi -\nabla^\mu J^i_\nu \right) =0,
\end{align}
where $T_{\mu\nu}$ is the matter stress-energy tensor. See Appendix~\ref{App:explEqs} for the definition of $\mathcal{G}^{i}_{\mu\nu}$, $P^i_\phi$ and $J^i_\nu$ and for the explicit form of Eq.~\eqref{HornEq2}.
%%%

\subsection{GR as a solution of Horndeski gravity}

We are interested in theories in which the scalar exhibits a tachyonic instability around solutions of GR. Hence we need to impose that the theory actually admits as a solution any spacetime of GR with $\phi=\phi_0=const$. This requires imposing certain conditions on the $G^{i}$ functions. These conditions have been fully worked out for shift-symmetric classes~\cite{Saravani:2019xwx} but not for theories that do not respect shift symmetry (and hence, can have a bare or effective mass).

The obvious thing one can do to do away with theories that do not admit solutions $\phi=\phi_0=const$, or $X=0$,  is to require that the $G_i$ functions  be analytic around $X=0$. In this case one can expand them in a power series in terms of $X$,
\begin{equation}
\label{G2analytical}
G_i=g_{i0}(\phi)+g_{i1}(\phi)X+\dots.
\end{equation}
However, imposing analyticity for the $G_i$ function is too restrictive, as we already know of a class of theories that can admit solutions with $\phi=\phi_0$ and have nonanalytic $G_i$ functions at $X=0$.\footnote{Another possibility is to extrapolate theories within Horndeski which admit GR as a solution only at leading order~\cite{McManus:2016kxu}. However this goes beyond the scope of this paper.}  The action of scalar Gauss-Bonnet (sGB) gravity contains a term of the form $\xi(\phi)\mathscr{G}$, where $\xi(\phi)$ is a generic function of $\phi$ and $\mathscr{G} =R^2 -4R_{\mu\nu} R^{\mu\nu} +R_{\mu\nu\rho\sigma}R^{\mu\nu\rho\sigma}$ is the GB invariant. One can retrieve this nonminimal coupling contribution from the action~\eqref{Horndeski}, with the following choice of the free functions~\cite{Kobayashi:2011nu},
\begin{equation}
\begin{aligned}
\label{G_GB}
G^{\note{GB}}_2= &8\xi^{(4)}X^2(3-\text{ln} X), \\
G^{\note{GB}}_3= &4\xi^{(3)}X(7-3\text{ln} X), \\
G^{\note{GB}}_4= &4\xi^{(2)}X(2-\text{ln} X), \\
G^{\note{GB}}_5= &-4\xi^{(1)}\text{ln} X,
\end{aligned}
\end{equation}
where $\xi^{(n)}\equiv\partial^n\xi/\partial\phi^n$. In the Horndeski representation the $G_i$ functions are nonanalytic in $X$ but there is an analytic representation of the action and the equations are analytic at $X=0$.  We stress that the case where $\xi=\phi$ and all the other free functions are shift symmetric is special, as Minkowski space is the only admissible GR solution~\cite{Saravani:2019xwx}.

sGB gravity is already known to exhibit spontaneous scalarization~\cite{Silva:2017uqg,Doneva:2017bvd}; hence, we should certainly relax our analyticity assumption on the $G_i$ functions in order to accommodate it. To this end, we rewrite the $G_i$ functions as a sum of an explicitly analytic part, which we label as $\tilde{G}_i$, and a nonanalytic part, coming from Eqs.~\eqref{G_GB}. Explicitly we have
\begin{gather}
\label{Gfunctions}
G_i(\phi,X)=\tilde{G}_i(\phi,X) + G^{\note{GB}}_i(\phi,X), \\
\label{eq:GtildeExp}
\tilde{G}_i(\phi,X)=g_{i0}(\phi)+g_{i1}(\phi)X+\dots
\end{gather}
where in Eq.~\eqref{eq:GtildeExp} we expanded $\tilde{G}_i$ as in Eq.~\eqref{G2analytical}.

The results and classification of Ref.~\cite{Saravani:2019xwx} for shift-symmetric theories suggest that $G_i$ that contain $\sqrt{|X|}$ might be another form of mild nonanalyticity that is compatible with GR solutions. However, we do not explore this possibility further. Moreover, in principle, there could be another type of nonshift-symmetric theories described by nonanalytic $G_i$ functions that admit all of the solutions of GR. This deserves further investigation, but we do not pursue it in this paper.

Once one has imposed the above conditions on the $G_i$ functions, the terms $-g_{30}(\phi)\DAlembert\phi$ and $g_{21}(\phi)X$ appear in the action and they coincide up to total derivative. Thus, without loss of generality we can set $g_{30}(\phi)=0$, which is equivalent to the redefinition $g_{21}(\phi)\rightarrow g_{21}(\phi)+2g_{30\phi}(\phi)$. Moreover, \textit{at the level of the linearized equations}, which is our interest here, the terms $g_{41}(\phi)$ and $-g_{50\phi}(\phi)$ give the same contribution. Hence, we similarly redefine $g_{41}(\phi)\rightarrow g_{41}(\phi)+g_{50\phi}(\phi)$.
%%%%

Let us now look explicitly at the equations of motion. The metric satisfies Einstein equations~\eqref{HornEq1}, which for any constant scalar field $\phi=\phi_0$ read
\begin{equation}
\label{EinstEq}
R_{\mu\nu}-\frac{1}{2}g_{\mu\nu}R+\Lambda g_{\mu\nu}=\tilde{\kappa}T_{\mu\nu},
\end{equation}
where
\begin{equation}\label{eq:Lambda,Kappa}
\Lambda=-g^0_{20}/2g^0_{40}, \qquad \tilde{\kappa}=\kappa/g^0_{40},
\end{equation}
provided that $g_{40}^0\neq0$. The superscript $0$ in $g^0_{20}$, $g^0_{40}$, etc., means that the function is evaluated at $\phi=\phi_0$.
The equations above imply clearly that the metric is a solution of GR Einstein equations and that all solutions of Einstein's equation are admissible.

Let us now take the scalar field equation~\eqref{HornEq2}, with the choice of functions of Eq.~\eqref{Gfunctions}. We keep only the terms which contain up to one derivative operator (which, as we see, can be only a second order operator) acting on $\phi$. With this choice we capture all the terms that contribute to the linearization of the equation around the constant value $\phi_0$ made in the next paragraph. We stress that first order derivatives do not contribute to the linearized equations. Indeed, these terms appear at least in the form $\nabla\phi\nabla\phi$, which, upon linearization, vanishes when the background field is constant. With this prescription, the scalar field equation takes the form
\begin{equation}
\label{ScalEq}
\tilde{g}^{\mu\nu}\nabla_\mu\nabla_\nu\phi +\frac{g_{20\phi}+g_{40\phi}R+\xi^{(1)}\mathscr{G}}{A(\phi)}=0,
\end{equation}
where
\begin{equation}\label{eq:Acoeff}
A(\phi)=g_{21}+g_{41}R,
\end{equation}
and the effective metric reads\footnote{Note that the effective metric~\eqref{effMetr} must have a Lorentzian signature in order for the linearized equation to be hyperbolic and hence describe the time evolution of the system. This imposes some further conditions on $g_{21}$ and $g_{41}$. In this paper, we implicitly assume that such conditions are  satisfied.}
\begin{equation}
\label{effMetr}
\tilde{g}^{\mu\nu}=g^{\mu\nu} - \frac{2g_{41}R^{\mu\nu}}{A(\phi)}.
\end{equation}

We now impose that $\phi=\phi_0$ is a solution of Eq.~\eqref{ScalEq}. There are two distinct cases for which this happens,
\begin{align}
\text{case I:}  \qquad & g^0_{20\phi}+g^0_{40\phi}R+\xi^{(1)}_0\mathscr{G} =0, \notag \\
\label{GRcond1}
& A_0\,\, \text{finite} ; \\
\text{case II:} \qquad & g^0_{20\phi}+g^0_{40\phi}R+\xi^{(1)}_0\mathscr{G} \neq0, \notag \\
\label{GRcond2}
& A_0\rightarrow \infty,
\end{align}
where
\begin{equation}\label{eq:A0}
A_0\equiv A(\phi_0)=g_{21}^0+g_{41}^0 R.
\end{equation}
Case II is rather interesting, as it provides a way to have a GR solution even when the term $g_{20\phi}+g_{40\phi}R+\xi^{(1)}\mathscr{G}$ does not depend on $\phi$ at all (or equivalently when $g^0_{20\phi\phi}=g^0_{40\phi\phi}=\xi^{(2)}_0=0$) and would otherwise act as a source term for the scalar field.
For example, as we see in more detail below, standard scalar-tensor theories belong to case II, as they correspond to $g_{40}=\phi$, $g_{41}=0$. They admit GR solutions only when $g_{21}(\phi)=2\omega(\phi)/\phi\rightarrow\infty$ for $\phi\rightarrow\phi_0$. Another interesting term in this context is that with $\xi=\phi$. As already mentioned, this choice leads to the $\phi\,\mathscr{G}$ term, which is shift symmetric, and the GB invariant would appear in the scalar field equation as a pure source for the scalar field. Thus, only theories that satisfy condition~\eqref{GRcond2} can afford to include this term and still admit GR solution. This possibility is absent in shift-symmetric theories~\cite{Saravani:2019xwx}.

Note that an analysis similar to the one presented here has been conducted in Ref.~\cite{Motohashi:2018wdq} for multiscalar-tensor theories, but with more restrictive assumptions that appear to exclude case II.

\subsection{Linearized scalar field equations}\label{Sec2}

Linearizing Eq.~\eqref{HornEq2} divided by $A(\phi)$ [or equivalently Eq.~\eqref{ScalEq}] for small $\delta\phi=\phi-\phi_0$ yields
\begin{equation}
\label{LinEq}
\tilde{g}^{\mu\nu}\nabla_\mu\nabla_\nu\delta\phi-m_\note{I}^2\delta\phi-m_\note{II}^2\delta\phi=0,
\end{equation}
where
\begin{align}
\label{massI}
m^2_\note{I} &=-\frac{g^0_{20\phi\phi}+g^0_{40\phi\phi}R+\xi^{(2)}_0\mathscr{G}}{A_0}, \\
\label{massII}
m^2_\note{II} &=\frac{g^0_{20\phi}+g^0_{40\phi}R+\xi^{(1)}_0\mathscr{G}}{A_0^2}\frac{\partial A}{\partial\phi}\bigg|_{\phi_0}
\end{align}
are the effective masses obtained in the two separate cases. We notice that the two cases give mutually exclusive contributions to the mass. Indeed, if relation~\eqref{GRcond1} holds, then $m_\note{II}=0$; and when the condition~\eqref{GRcond2} holds, $m_\note{I}$ vanishes. Note that in the latter case,  $A_0\to \infty$ and  having a nonzero effective mass $m_\note{II}$ requires that $\frac{\partial A}{\partial\phi}\big|_{\phi_0} \to \infty$ is such that
\begin{equation}\label{eq:A0primecondition}
\frac{1}{A_0^2}\frac{\partial A}{\partial\phi}\bigg|_{\phi_0}\neq 0\quad \text{and finite}.
\end{equation}
Hence, around $\phi=\phi_0$ it must be $A(\phi)\sim (\phi-\phi_0)^{-1}$.

We can now single out the theories which can exhibit a tachyonic instability around a GR background. They either satisfy condition~\eqref{GRcond1} and have $m^2_\note{I}<0$  or they satisfy condition~\eqref{GRcond2} and have $m^2_\note{II}<0$.
We stress that our perturbative analysis is done around  a GR background and we perturb only the scalar without taking into account its backreaction to the metric. This approximation (decoupling) offers drastic simplification. Though doing a full analysis of perturbation that includes the metric might be necessary for quantitative estimates of the thresholds associated with the tachyonic instability, we consider our approximation to be adequate for the more qualitative task of identifying theories that exhibit the instability.

\section{Theories with tachyonic instability}\label{Sec3}
\subsection{The minimal actions}
We now analyze what the theories are that belong in one of the categories we identified above.  At first, we write down for each case the minimal action that consists of all the terms that contribute to the linearized equation and admit GR solutions when $\phi=\phi_0$. Let us redefine the scalar field such that $\phi_0=0$. The minimal action for case I is
\begin{multline}\label{eq:ActionCaseI}
S_\note{I}=\int\dd^4x\frac{\sqrt{-g}}{2\kappa}\bigg[R-2\Lambda+(a_{21}+a_{41}R)X \\ +a_{41}R_{\mu\nu}\nabla^\mu\phi\nabla^\nu\phi 
 -\frac{m_\phi^2\phi^2-\alpha\phi^2 R-\beta\,\phi^2\,\mathscr{G}}{2}\bigg]+S_\note{M},
\end{multline}
whereas for case II we have
\begin{multline}\label{eq:ActionCaseII}
S_\note{II}=\int\dd^4x\frac{\sqrt{-g}}{2\kappa}\bigg[R-2\Lambda+\frac{b_{21}+b_{41}R}{\phi}X  \\
+\frac{b_{41}}{\phi}R_{\mu\nu}\nabla^\mu\phi\nabla^\nu\phi +\tau\phi +\eta\,\phi R +\lambda\,\phi\,\mathscr{G}\bigg] + S_\note{M}.
\end{multline}
We normalized the actions~\eqref{eq:ActionCaseI} and~\eqref{eq:ActionCaseII} by the constant multiplying $R$, which is equivalent to setting $g_{40}^0=1$. Moreover, we can identify the constants written in the actions~\eqref{eq:ActionCaseI} and~\eqref{eq:ActionCaseII} in terms of the function $g_{ij}$ evaluated at $\phi=0$,
\begin{equation}\label{eq:constantsGs}
\begin{aligned}
\Lambda=-\frac{g_{20}^0}{2}, & \qquad \tau=g_{20\phi}^0, \qquad m_\phi^2=-g_{20\phi\phi}^0, \\
a_{21}=(\phi\,g_{21})_\phi^0, & \qquad b_{21}=\left(\phi^2\,g_{21}\right)_\phi^0, \\
                    & \qquad \eta=g_{40\phi}^0, \qquad \alpha=g_{40\phi\phi}^0, \\
a_{41}=(\phi\,g_{41})_\phi^0, & \qquad b_{41}=\left(\phi^2\,g_{41}\right)_\phi^0, \\
                    & \qquad \lambda=\xi^{(1)}_0, \qquad \beta=\xi^{(2)}_0.
\end{aligned}
\end{equation}
The actions above could be supplemented with any term that does not contribute to the linearized equations without affecting the onset of the tachyonic instability. However, such nonlinear terms are crucial for determining the end state of the instability and the properties of scalarized solutions~\cite{Silva:2018qhn,Macedo:2019sem}. Hence, one can start from the minimal models above and bootstrap their way to theories that exhibit scalarization bur differ quantitatively thanks to terms that introduce different nonlinear corrections.

\subsection{Equivalence between case I and case II}

So far we have treated case I and case II separately because they lead to distinct contributions to the effective mass and, naively, they appear to be qualitatively different. Actually, they are equivalent as different representations of the same physics. Indeed, one can start from action~\eqref{eq:ActionCaseII}, perform the scalar field redefinition
\begin{equation}\label{eq:FieldRescalingLinear}
\phi \rightarrow \phi^2\,,
\end{equation}
and obtain action \eqref{eq:ActionCaseI} with the correspondence of parameters
\begin{equation}\label{eq:ConstantsRedefinition}
\begin{split}
a_{21}=4\,b_{21}, \qquad a_{41}=4\,b_{41}, \\
m_\phi^2= - 2\,\tau, \qquad \alpha= 2\,\eta, \qquad \beta= 2\,\lambda.
\end{split}
\end{equation}
Hence, any theory in the minimal action of case II can be mapped onto an equivalent case I theory, at least in what regards their linear behavior and the onset of the tachyonic instability.

This observation simplifies our analysis and reduces significantly the different scenarios of scalarization.

\subsection{Models of scalarization}

Having shown that the two cases are equivalent, we now focus on the action outlined in Eq.~\eqref{eq:ActionCaseI} and  consider each term that contributes to the mass separately. This helps us identify its relation with known models of scalarization.  The term that contains $X$ in the action~\eqref{eq:ActionCaseI} contributes to the effective mass only as a multiplicative constant on a GR background. $a_{21}$ can be set to $1$ through a constant rescaling of the scalar and we do so implicitly in what follows. The  $a_{41}$  is rather distinct from the rest so, for the time being, let us set $a_{41}=0$ and reduce the $X$-dependent term  to the canonical kinetic term. We relax this assumption in the next section.

The first term that contributes to the effective mass is the bare mass of the scalar field $m_\phi^2$. If the mass square is negative, it could lead to a tachyonic instability that would persist in flat space. So, we disregard this possibility. If it is positive, it needs to be sufficiently small not to prohibit the other terms from inducing a tachyonic instability. A small bare mass can actually be beneficial, as it can help suppress the non-GR effects away from the compact object. One can generalize the bare mass term to a full-fledged potential and this would introduce nonlinearities that could affect the end point of scalarization~\cite{Macedo:2019sem}. However, it is rather clear that a bare mass term or a potential cannot lead to scalarization.

Next we consider the coupling term between $\phi$ and the GB invariant. For the choice $m_\phi=\alpha=0$ (and $a_{21}=1$, $a_{41}=0$) one has the action
\begin{equation}\label{eq:CaseI_GBTerm_Action_0}
S=\int \dd^4x \frac{\sqrt{-g}}{2\kappa}\left[R-\frac{1}{2}\nabla^\mu \phi \nabla_\mu \phi+\beta \phi^2\mathscr{G}\right]+S_{\note{M}},
\end{equation}
This is the quadratic coupling scalarization model considered in Ref.~\cite{Silva:2017uqg}.
Allowing for a more general coupling function one gets the action considered in Refs.~\cite{Doneva:2017bvd,Silva:2017uqg},
\begin{equation}\label{eq:CaseI_GBTerm_Action}
S=\int \dd^4x \frac{\sqrt{-g}}{2\kappa}\left[R-\frac{1}{2}\nabla^\mu \phi \nabla_\mu \phi+\xi(\phi)\mathscr{G}\right]+S_{\note{M}},
\end{equation}
where, from the condition~\eqref{GRcond1} one can infer that $\xi_\phi(0)=0$. This condition guarantees that the leading term in $\xi(\phi)$ is indeed $\phi^2$.

Finally, if we set $m_\phi=\beta=0$, we have
\begin{equation}\label{eq:CaseI_RTerm_Action_0}
S=\int \dd^4x \frac{\sqrt{-g}}{2\kappa}\left[\left(1+\frac{\alpha\phi^2}{2}\right)R-\frac{1}{2}\nabla^\mu \phi \nabla_\mu \phi \right]+S_{\note{M}}.
\end{equation}
We can generalize this theory in a similar fashion as above and write
\begin{equation}\label{eq:CaseI_RTerm_Action}
S=\int \dd^4x \frac{\sqrt{-g}}{2\kappa}\left[f(\phi)R-\frac{1}{2}\nabla^\mu \phi \nabla_\mu \phi\right]+S_{\note{M}},
\end{equation}
where we assume $f(0)\neq0$.  The condition~\eqref{GRcond1} implies  $f_\phi(0)=0$ , and $f_{\phi\phi}(0)<0$ is the requirement for a tachyonic instability of the theory.

One may be tempted to think that this is a new model. However, we recall that we can always perform a redefinition of the scalar field, as we did to relate the minimal actions of case I and case II. Indeed, consider the redefinition
\begin{equation}\label{eq:FieldRescaling}
\Phi=f(\phi)\,.
\end{equation}
Action \eqref{eq:CaseI_RTerm_Action} can be rewritten as
\begin{equation}\label{eq:CaseII_STT}
S=\int \dd^4 x \frac{\sqrt{-g}}{2\kappa}\left[\Phi R-\frac{\omega(\Phi)}{\Phi} \nabla^\mu \Phi \nabla_\mu \Phi\right] + S_\note{M}\,,
\end{equation}
if we just introduce the definition
\begin{equation}\label{eq:RescalingSTT}
\omega(\Phi)\equiv\frac{\Phi}{2f'^2(\phi)}.
\end{equation}
Action~\eqref{eq:CaseII_STT} is that of scalar-tensor theories written in the so-called Jordan frame (see e.g.~\cite{Faraoni:2004pi}). The condition $f_\phi(0)=0$ translates into $\omega(\Phi_0)\to \infty$, where $\Phi_0=f(0)$. This picks a specific subclass of scalar-tensor theories, which is precisely that originally considered by DEF~\cite{Damour:1993hw}.\footnote{Albeit it is usually studied in the Einstein frame, obtained by a conformal transformation of the metric and suitably redefining the scalar field.}

Indeed, the minimal model in action~\eqref{eq:CaseI_RTerm_Action_0} corresponds to $f(\phi)=1+\alpha \phi^2/2$ and hence $\Phi=1+\alpha\phi^2/2$,
\begin{equation}
\omega(\Phi)=\frac{\Phi}{4\alpha(\Phi-1)}=\frac{1}{4\alpha}+\frac{1}{4\alpha(\Phi-1)}\,,
\end{equation}
and $\Phi_0=1$. One can easily verify that the most commonly studied DEF model corresponds in the Jordan frame to
\begin{equation}\label{eq:omega}
\omega_\note{DEF}(\Phi)=-\frac{3}{2}-\frac{1}{2\beta_\note{DEF} \log\Phi}\,,
\end{equation}
where we have used the subscript DEF to distinguish the commonly used $\beta$ parameter from our notation above. As $\Phi\to \Phi_0=1$ one has
\begin{equation}
\omega_\note{DEF}(\Phi)\to -\frac{1}{2\beta_\note{DEF}(\Phi-1)}\,,
\end{equation}
which is precisely the same behavior as our minimal model up to a redefinition of constants. The two models are indistinguishable at the linear level.

We close this section with a few remarks. First, the scalar field redefinition that related the $f(\phi)R$ model with the DEF class was basically mapping a case I theory onto a case II theory. Indeed, one can straightforwardly identify the DEF class as a subcase of the action~\eqref{eq:ActionCaseII}, with the constant coefficients generalized to functions of $\phi$.  Secondly, these results clearly show that some models that might appear as new are simply combinations of known models rewritten after a scalar field redefinition. For instance, the action
\begin{equation}\label{eq:Action_R_GB}
S=\int\dd^4 x \frac{\sqrt{-g}}{2\kappa}\left[\phi R +2\frac{\omega(\phi)}{\phi}X+\eta\phi\,\mathscr{G}\right]+S_\note{M},
\end{equation}
with the condition $\omega(\phi_0)\to \infty$ for some $\phi_0$ would yield a seemingly intriguing case II model upon linearization, but it can straightforwardly be mapped onto a combination of actions~\eqref{eq:CaseI_GBTerm_Action} and~\eqref{eq:CaseI_RTerm_Action}.
%%%%%%

\subsection{Disformal transformations and matter coupling}\label{Sec5}

Throughout the paper we have assumed that the matter couples minimally to the metric only. Moreover, in the previous section we had set the coefficients $a_{41}$ of the action~\eqref{eq:ActionCaseI} to the specific value $a_{41}=0$. At linear level (which is our main interest throughout), it turns out that one can always do so without loss of generality by relaxing the matter coupling assumption.

To show this, let us start with action~\eqref{eq:ActionCaseI}  and elevate all of the constants to generic functions of $\phi$ (retaining the minimal coupling to matter, described by some generic fields $\Psi^A$),
\begin{multline}\label{eq:GeneralAction}
S=\int\dd^4 x\frac{\sqrt{-g}}{2\kappa}\Big[(g_{40}(\phi)+g_{41}(\phi)X)R\\
 +g_{41}(\phi)R_{\mu\nu}\nabla^\mu\phi\nabla^\nu\phi
 +g_{21}(\phi)X \\
+g_{20}(\phi)+\xi(\phi)\mathscr{G}\Big] +S_\note{M}\left[g_{\mu\nu},\Psi^A\right].
\end{multline}
We stress that the unknown functions of $\phi$ are assumed to be such that linearizing this action around $\phi=0$ must yield~\eqref{eq:ActionCaseI}, with the identification of the constants~\eqref{eq:constantsGs}. Consider now a disformal transformation of the form
\begin{equation}\label{eq:disformal}
g_{\mu\nu} \rightarrow C(\phi)\left[g_{\mu\nu}+D(\phi)\nabla_\mu\phi\nabla_\nu\phi\right].
\end{equation}
This transformation leaves  the Horndeski action~\eqref{Horndeski} formally invariant~\cite{Bettoni:2013diz,Zumalacarregui:2013pma}. A transformation with $D=0$ is called \textit{ conformal}, whereas for $C=1$ one has a \textit{purely disformal} transformation.
Applying this transformation to \eqref{eq:GeneralAction} and keeping only the terms which contribute to the linear level in the equations yields
\begin{multline}\label{eq:DisfGeneralAction}
S=\int \dd^4 x \frac{\sqrt{-g}}{2\kappa} \Big[(\bar{g}_{40}(\phi)+\bar{g}_{41}(\phi)X)R \\
+\bar{g}_{41}(\phi)R_{\mu\nu}\nabla^\mu\phi\nabla^\nu\phi
 +\bar{g}_{21}(\phi)X+\bar{g}_{20}(\phi)+\xi(\phi)\,\mathscr{G} \Big] \\ +S_{\note{M}}\left[C(\phi)\left(g_{\mu\nu}+D(\phi)\nabla_\mu\phi\nabla_\nu\phi\right), \Psi^A\right],
\end{multline}
where we made explicit the disformal coupling in the matter sector, and the new functions are defined as follows,~\footnote{We derived independently the effect of the disformal transformation~\eqref{eq:disformal} on the Horndeski Lagrangian~\eqref{Horndeski}. However, there is a mismatch with the results of~\cite{Bettoni:2013diz}. See Appendix~\ref{App:Disftransf}.}
\begin{align}\label{eq:FunctionsDisf}
  \bar{g}_{20} = & \, C^2 g_{20}\\ \label{eq:g21}
  \bar{g}_{21} = & \, C g_{21}-C^2D g_{20}-3g_{40}\frac{C_\phi^2}{C}-6g_{40\phi}C_{\phi} \\ \label{eq:g40}
  \bar{g}_{40} = & \, Cg_{40}, \\ \label{eq:g41}
  \bar{g}_{41} = & \, g_{41}-CDg_{40}-4\frac{C_\phi}{C}\xi^{(1)},
\end{align}
whereas $\xi(\phi)$ remains invariant. Here we are using again the same convention that a subscript $\phi$ denotes a derivative with respect to $\phi$. Hence, the action~\eqref{eq:DisfGeneralAction} yields field equations whose linear perturbation is formally invariant under the transformation~\eqref{eq:disformal}.

One notices that two out of the five functions $g_{40}$, $g_{41}$, $g_{21}$, $C$ and $D$ are redundant. That is, one can always perform a disformal transformation and choose $C$ and $D$ in order to redefine two of $g_{40}$, $g_{41}$, $g_{21}$.
For example, from Eq.~\eqref{eq:g41} one can set $\bar{g}_{41}=0$, by choosing
\begin{equation}\label{eq:DisfChoice}
D=\frac{g_{41}}{Cg_{40}}-4\frac{C_{\phi}}{C^2g_{40}}\xi^{(1)}.
\end{equation}
This choice fixes uniquely the disformal function $D$. This implies that the condition $g_{41}=0$ imposed throughout the previous section is equivalent to a specific type of disformal coupling. In other words, though having a nonzero $a_{41}$ does lead to a new theory, this theory is simply one of the known scalarization models, or a combination thereof, disformally coupled to matter (see~\cite{Minamitsuji:2016hkk} for a discussion of DEF spontaneous scalarization plus a disformal coupling).

For example, let us indeed impose Eq.~\eqref{eq:DisfChoice} in order to set $\bar{g}_{41}=0$ and we further choose
\begin{equation}\label{eq:ConfChoice}
C(\phi)=\frac{1}{g_{40}(\phi)},
\end{equation}
and redefine the scalar field as
\begin{equation}\label{eq:RescChoice}
\varphi = \varphi(\phi), \qquad \varphi'(\phi) = \frac{\sqrt{\bar{g}_{21}(\phi)}}{2},
\end{equation}
where $\bar{g}_{21}(\phi)$ is defined in Eq.~\eqref{eq:g21}.
With these choices,  action~\eqref{eq:DisfGeneralAction} takes the form
\begin{multline}
\label{eq:DisfGeneralActionEF}
S_\note{E}=\frac{1}{2\kappa}\int \dd^4 x \sqrt{-g} \left[R+V(\varphi)-2\partial_\mu\varphi\,\partial^\mu\varphi+F(\varphi)\mathscr{G} \right] \\ +S_{\note{M}}\left[G(\varphi)\left( g_{\mu\nu}+H(\varphi)\nabla_\mu\varphi\nabla_\nu\varphi \right), \Psi^A\right],
\end{multline}
where we defined the new functions
\begin{equation}
\begin{split}
V(\varphi)=\bar{g}_{20}(\phi(\varphi)), \qquad F(\varphi)=\xi(\phi(\varphi)), \\
G(\varphi)=C(\phi(\varphi)), \qquad H(\varphi) = \frac{4D(\phi(\varphi))}{\bar{g}_{21}(\phi(\varphi))}.
\end{split}
\end{equation}
For $\xi(\phi)=0$, this action reduces to the spontaneous scalarization model with disformal coupling studied for the first time in~\cite{Minamitsuji:2016hkk}.

\section{Discussion}\label{Sec6}

We have considered the Horndeski action and tried to identify classes of theories within it that exhibit spontaneous scalarizaton triggered by a tachyonic instability. We first determined the conditions that need to be satisfied so that solutions of general relativity are admissible. We probed whether or not there will be a tachyonic instability by calculating the effective mass of scalar perturbation on a {\em fixed} spacetime background that is a solution of Einstein's equations. Though this approximation neglects backreaction, we consider it adequate for simply identifying scalarization models.

Our analysis allowed us to determine a minimal action that contains all of the terms that contribute to the effective mass at linearized level.
This can be thought of as containing four distinct terms that contribute to scalarization. Through suitable field redefinitions, one of them can be directly linked to the known DEF model~\cite{Damour:1993hw} and another to the scalar-Gauss-Bonnet scalarization models~\cite{Doneva:2017bvd,Silva:2017uqg}. The third term can be thought of as a disformal coupling to matter and relates to a model studied in Ref.~\cite{Minamitsuji:2016hkk}. The fourth term comes from a potential for a scalar and, although it cannot trigger spontaneous scalarization on its own, it affects the onset of the tachyonic instability in all other models.

One can start from our minimal action, supplement it with terms that contribute only nonlinearly to the scalar equation, and construct scalarization models. The onset of the tachyonic instability that will kickstart scalarization will be determined by the minimal action, while the end state depends on the choice of the extra term that contributes nonlinearly. This is because scalarization is triggered by a linear tachyonic instability and later quenched by nonlinear effects.
We leave the study of the strong field phenomenology of such models for future work.

\section*{ACKNOWLEDGEMENTS}

We thank Antoine Lehébel for pointing out to us the omission of a term in some of our actions in an earlier version of this manuscript. T. P. S. acknowledges partial support from the STFC Consolidated Grant No. ST/P000703/1. We also acknowledge networking support by the COST Action GWverse Grant No. CA16104.

%%%%%
\appendix
\section{Horndeski equations of motion}\label{App:explEqs}
We give here  explicit expressions for the terms in the field equations presented in Sec.~\ref{Sec1}. Throughout the  appendix we use the notation $\phi_\mu\equiv \nabla_\mu \phi$ and $\phi_{\mu\nu}\equiv \nabla_\mu\nabla_\nu\phi$. The $\mathcal{G}^{i}_{\mu\nu}$ functions appearing in the modified Einstein equations are
\begin{widetext}
\begin{subequations}
\begin{equation}
\label{G2}
\mathcal{G}^2_{\mu\nu}=-\frac{1}{2}G_{2X}\phi_\mu\phi_\nu-\frac{1}{2}G_2 g_{\mu\nu}
\end{equation}
\begin{equation}
\label{G3}
\mathcal{G}^3_{\mu\nu}=\frac{1}{2}G_{3X}\DAlembert\phi\phi_\mu\phi_\nu+\nabla_{(\mu} G_3\phi_{\nu)}-\frac{1}{2} g_{\mu\nu}\nabla_\lambda G_3\phi^\lambda
\end{equation}
\begin{equation}
\label{G4}
\begin{split}
\mathcal{G}^4_{\mu\nu}= & \, G_4G_{\mu\nu}-\frac{1}{2}G_{4X}R\phi_\mu\phi_\nu -\frac{1}{2}G_{4XX}\left[(\DAlembert\phi)^2-(\phi_{\alpha\beta})^2 \right]\phi_\mu\phi_\nu-G_{4X}\DAlembert\phi\phi_{\mu\nu} \\
+ & \, G_{4X}\phi_{\mu\lambda}\phi^\lambda_{\;\nu}+2\nabla_\lambda G_{4X}\phi^\lambda_{\;(\mu}\phi_{\nu)} -\nabla_\lambda G_{4X}\phi^\lambda\phi_{\mu\nu} +g_{\mu\nu}(G_{4\phi}\DAlembert\phi-2XG_{4\phi\phi})\\
+ & \, g_{\mu\nu}\big\lbrace -2G_{4\phi X}\phi_{\alpha\beta}\phi^\alpha\phi^\beta +G_{4XX}\phi_{\alpha\lambda}\phi^\lambda_{\;\beta}\phi^\alpha\phi^\beta +\frac{1}{2}G_{4X}\left[ (\DAlembert\phi)^2-(\phi_{\alpha\beta})^2 \right]  \big\rbrace \\
+ & \, 2\big[ G_{4X}R_{\lambda (\mu}\phi_{\nu )}\phi^\lambda -\nabla_{(\mu}G_{4X}\phi_{\nu)}\DAlembert\phi\big]-g_{\mu\nu}\left[ G_{4X}R^{\alpha\beta}\phi_\alpha\phi_\beta-\nabla_\lambda G_{4X}\phi^\lambda\DAlembert\phi\right] \\
+ & \, G_{4X}R_{\mu\alpha\nu\beta}\phi^\alpha\phi^\beta -G_{4\phi}\phi_{\mu\nu}-G_{4\phi\phi}\phi_\mu\phi_\nu +2G_{4\phi X}\phi^\lambda\phi_{\lambda(\mu}\phi_{\nu )}
-G_{4XX}\phi^\alpha\phi_{\alpha\mu}\phi^\beta\phi_{\beta\nu}
\end{split}
\end{equation}
\begin{equation}
\label{G5}
\begin{split}
\mathcal{G}^5_{\mu\nu}= & \, G_{5X}R_{\alpha\beta}\phi^\alpha\phi^\beta_{\;(\mu}\phi_{\nu)} -G_{5X}R_{\alpha(\mu}\phi_{\nu)}\phi^\alpha\DAlembert\phi  -\frac{1}{2}G_{5X}R_{\alpha\beta}\phi^\alpha\phi^\beta\phi_{\mu\nu} -\frac{1}{2}G_{5X}R_{\mu\alpha\nu\beta}\phi^\alpha\phi^\beta\DAlembert\phi\\
+ & \, G_{5X}R_{\alpha\lambda\beta(\mu}\phi_{\nu)}\phi^\lambda\phi^{\alpha\beta} +G_{5X}R_{\alpha\lambda\beta(\mu}\phi_{\nu)}^\lambda\phi^\alpha\phi^\beta -\frac{1}{2}\left\{\nabla_{(\mu}[G_{5X}\phi^\alpha]\phi_{\alpha\nu)} -\nabla_{(\mu}[G_{5X}\phi_{\nu)}]\right\}\DAlembert\phi\\
- & \, \nabla_\lambda[G_{5\phi}\phi_{(\mu}]\phi_{\nu)}^{\;\lambda} +\frac{1}{2}\left[ \nabla_\lambda(G_{5\phi}\phi^\lambda) -\nabla_\alpha(G_{5X}\phi_\beta)\phi^{\alpha\beta} \right]\phi_{\mu\nu}
+\nabla^\alpha G_5\phi^\beta R_{\alpha(\mu\nu)\beta} -\nabla_{(\mu}G_5G_{\nu)\lambda}\phi^\lambda \\
+ & \, \frac{1}{2}\nabla_{(\mu}G_{5X}\phi_{\nu)}\left[ (\DAlembert\phi)^2-(\phi_{\alpha\beta})^2 \right]-\nabla^\lambda G_5 R_{\lambda(\mu}\phi_{\nu)} + \nabla_\alpha[G_{5X}\phi_\beta]\phi^\alpha_{\;(\mu}\phi_{\nu)}^{\;\beta} -\frac{1}{2}G_{5X}G_{\alpha\beta}\phi^{\alpha\beta}\phi_\mu\phi_\nu  \\
- & \, \nabla_\beta G_{5X}\left[ \DAlembert\phi\phi^\beta_{\;(\mu}-\phi^{\alpha\beta}\phi_{\alpha(\mu} \right] \phi_{\nu)}  +\frac{1}{2}\phi^\alpha\nabla_\alpha G_{5X} \left[ \DAlembert\phi\phi_{\mu\nu}-\phi_{\beta\mu}\phi^\beta_{\;\nu} \right] -\frac{1}{2}G_{5X}\DAlembert\phi\phi_{\alpha\mu}\phi^\alpha_{\;\nu}\\
+ & \, \frac{1}{2}G_{5X}(\DAlembert\phi)^2\phi_{\mu\nu}+\frac{1}{12}G_{5XX}\left[(\DAlembert\phi)^3 -3\DAlembert\phi(\phi_{\alpha\beta})^2  +2(\phi_{\alpha\beta})^3\right]\phi_\mu\phi_\nu +\frac{1}{2}\nabla_\lambda G_5 G_{\mu\nu}\phi^\lambda\\
+ & \, g_{\mu\nu} \Biggl\{ -\frac{1}{6} G_{5X} \left[ (\DAlembert\phi)^3 -3\DAlembert\phi(\phi_{\alpha\beta})^2+2(\phi_{\alpha\beta})^3 \right] + \nabla_\alpha G_5 R^{\alpha\beta}\phi_\beta  -\frac{1}{2}\nabla_\alpha(G_{5\phi}\phi^\alpha)\DAlembert\phi \\
+ & \, \frac{1}{2}\nabla_\alpha(G_{5\phi}\phi_\beta)\phi^{\alpha\beta}-\frac{1}{2}\nabla_\alpha G_{5X}\nabla^\alpha X\DAlembert\phi  +\frac{1}{2} \nabla_\alpha G_{5X}\nabla_\beta X \phi^{\alpha\beta}-\frac{1}{4}\nabla^\lambda G_{5X}\phi_\lambda \left[ (\DAlembert\phi)^2-(\phi_{\alpha\beta})^2 \right]\\
+ & \, \frac{1}{2}G_{5X}R_{\alpha\beta}\phi^\alpha\phi^\beta\DAlembert\phi -\frac{1}{2}G_{5X}R_{\alpha\lambda\beta\rho}\phi^{\alpha\beta}\phi^\lambda\phi^\rho \Biggr\}. \\
\end{split}
\end{equation}
\end{subequations}
The function $P^i_\phi$ and $J^i_\mu$ appearing in the scalar field equations are
\begin{align}
\label{P2}
P^2_\phi & =G_{2\phi}, \\
\label{P3}
P^3_\phi & =\nabla_\mu G_{3\phi}\phi^\mu, \\
\label{P4}
P^4_\phi & =G_{4\phi}R+G_{4\phi X}\left[ (\DAlembert\phi)^2-(\phi_{\alpha\beta})^2 \right], \\
\label{P5}
P^5_\phi & =-\nabla_\mu G_{5\phi}G^{\mu\nu}\phi_\nu-\frac{1}{6}G_{5\phi X}\left[ (\DAlembert\phi)^3 -3\DAlembert\phi(\phi_{\alpha\beta})^2+2(\phi_{\alpha\beta})^3 \right],
\end{align}
\begin{align}
\label{J2}
J^2_\mu = & \,-\mathcal{L}_{2X}\phi_\mu, \\
\label{J3}
J^3_\mu = & \,-\mathcal{L}_{3X}\phi_\mu+G_{3X}\nabla_\mu X+2G_{3\phi}\phi_\mu, \\
\notag
J^4_\mu = & \,-\mathcal{L}_{4X}\phi_\mu+2G_{4X}R_{\mu\nu}\phi^\nu-2G_{4XX}(\DAlembert\phi\nabla_\mu X-\nabla^\nu X \phi_{\mu\nu})\\
\label{J4}
& \, -2G_{4\phi X}(\DAlembert\phi\phi_\mu+\nabla_\mu X), \\
\notag
J^5_\mu = & \, -\mathcal{L}_{5X}\phi_\mu-2G_{5\phi}G_{\mu\nu}\phi^\nu \\
\notag
& \, -G_{5X}\left[ G_{\mu\nu}\nabla^\nu X +R_{\mu\nu}\DAlembert\phi\phi^\nu-R_{\nu\lambda}\phi^\nu\phi^\lambda_{\;\mu} -R_{\alpha\mu\beta\nu}\phi^\nu\phi^{\alpha\beta}\right] \\
\notag
& \, +G_{5XX}\left\{ \frac{1}{2}\nabla_\mu X\left[ (\DAlembert\phi)^2-(\phi_{\alpha\beta})^2 \right]-\nabla_\nu X (\DAlembert\phi \phi_\mu^{\;\nu}-\phi_{\alpha\mu}\phi^{\alpha\nu}) \right\}\\
\label{J5}
& \, +G_{5\phi X} \left\{ \frac{1}{2}\phi_\mu\left[ (\DAlembert\phi)^2-(\phi_{\alpha\beta})^2 \right]+\DAlembert\phi\nabla_\mu X-\nabla^\nu X \phi_{\mu\nu} \right\}.
\end{align}
The explicit expression for the scalar field equation is
\begin{equation}
\begin{split}
&  -G_{2\phi} -G_{2X}\DAlembert\phi -G_{2\phi X} \phi^\mu \phi_\mu +G_{2XX}\phi^\mu\phi^\nu\phi_{\mu\nu}
+2G_{3\phi}\DAlembert\phi \\
& +G_{3X}\left[ (\DAlembert\phi)^2-R_{\mu\nu}\phi^\mu\phi^\nu - (\phi_{\mu\nu})^2 \right] + G_{3\phi\phi}\phi_\mu\phi^\mu   + G_{3\phi X}\phi^\mu \left( \phi_\mu \DAlembert\phi -2 \phi^\nu \phi_{\mu\nu} \right)\\
& + G_{3XX}\phi^\mu\phi^\nu \left( \tensor{\phi}{^\lambda_\mu}\phi_{\lambda\nu}-\phi_{\mu\nu}\DAlembert\phi \right)
-G_{4\phi}R +G_{4X}G_{\mu\nu}\phi^{\mu\nu}  \\
&  +G_{4\phi X} \left[ 4 R_{\mu\nu}\phi^\mu\phi^\nu-R\phi^\mu\phi_\mu-3 (\DAlembert\phi)^2+3(\phi_{\mu\nu})^2 \right] + G_{4XX} \big\{ \DAlembert\phi \left[ 3 (\phi_{\lambda\sigma})^2-(\DAlembert\phi)^2 \right] \\
& -2(\phi^{\mu\nu})^3 + \phi^\mu\phi^\nu \big( R\phi_{\mu\nu}-4R_{\mu\lambda}\tensor{\phi}{^\lambda_\nu} +2R_{\mu\nu}\DAlembert\phi-2R_{\mu\lambda\nu\sigma}\phi^{\lambda\sigma} \big) \big\}\\
& +2 G_{4\phi\phi X}\phi^\mu \left( \phi^\nu \phi_{\mu\nu}-\phi_\mu \DAlembert\phi \right) + G_{4\phi XX}\phi^\mu \big\{ 4\phi^\nu \left( \phi_{\mu\nu}\DAlembert\phi-\phi_{\lambda\nu}\tensor{\phi}{^\lambda_\mu} \right)\\
&-\phi_\mu \big[ (\DAlembert\phi)^2-(\phi_{\lambda\sigma})^2 \big] \big\} +G_{4XXX}\phi^\mu\phi^\nu \big\{ 2\tensor{\phi}{^\lambda_\mu}\left( \phi_{\lambda\sigma}\tensor{\phi}{_\nu^\sigma}- \phi_{\lambda\nu}\DAlembert\phi \right)\\
&  + \phi_{\mu\nu}\big[ (\DAlembert\phi)^2-(\phi_{\lambda\sigma})^2 \big] \big\}
 -2 G_{5\phi}G_{\mu\nu}\phi^{\mu\nu} +\frac{1}{2}G_{5X} \big[ R (\DAlembert\phi)^2+2R^{\mu\lambda}R_{\nu\lambda}\phi_\mu\phi^\nu\\
& -R_{\mu\nu}R\phi^\mu\phi^\nu+2R^{\lambda\sigma}R_{\mu\lambda\nu\sigma}\phi^\mu\phi^\nu -R^{\mu\lambda\sigma\rho}R_{\nu\lambda\sigma\rho}\phi_\mu\phi^\nu -R(\phi_{\mu\nu})^2 -4R^{\mu\nu}\phi_{\mu\nu}\DAlembert\phi \\
& + 4R^{\mu\nu}\phi_{\lambda\nu}\tensor{\phi}{^\lambda_\mu}+2R_{\mu\lambda\nu\sigma}\phi^{\mu\nu}\phi^{\sigma\lambda} \big] - G_{5\phi\phi}G_{\mu\nu}\phi^\mu\phi^\nu
+ G_{5\phi X} \big\{ \phi^\mu \phi^\nu \big[ 4R_{\mu\lambda}\tensor{\phi}{^\lambda_\nu}-2R_{\mu\nu}\DAlembert\phi\\
& -R\phi_{\mu\nu} +2R_{\mu\lambda\nu\sigma}\phi^{\lambda\sigma} \big] + \frac{2}{3} \big[ 2(\phi^{\mu\nu})^3+\DAlembert\phi \big( (\DAlembert\phi)^2-3(\phi_{\mu\nu})^2 \big)\big] -G_{\mu\nu}\phi^{\mu\nu}\phi^\lambda\phi_\lambda \big\}\\
& +\frac{1}{6}G_{5XX}\big\{ 3\phi^{\mu\nu}\phi_{\lambda\sigma}\big( \phi_{\mu\nu}\phi^{\lambda\sigma} -2\tensor{\phi}{^\lambda_\mu}\tensor{\phi}{^\sigma_\nu}\big) +\DAlembert\phi \big[ 8(\phi^{\mu\nu})^3\\
& +\DAlembert\phi \left( (\DAlembert\phi)^2-6(\phi_{\mu\nu})^2 \right) \big] -3\phi^\mu\phi^\nu \big[ 2R^{\lambda\sigma}\phi_{\lambda\mu}\phi_{\sigma\nu}-2G_{\lambda\sigma}\phi^{\lambda\sigma}\phi_{\mu\nu} +R_{\mu\nu}(\DAlembert\phi)^2 \\
& -\tensor{\phi}{^\lambda_\nu}\left( R\phi_{\lambda\mu}+4R_{\lambda\mu}\DAlembert\phi\right)+ 4 R_{\mu\lambda}\phi^{\lambda\sigma}\phi_{\sigma\nu}-R_{\mu\nu}(\phi_{\sigma\lambda})^2+2R_{\mu\sigma\nu\rho}\phi^{\sigma\lambda}\tensor{\phi}{^\rho_\lambda}-2R_{\mu\sigma\nu\rho}\phi^{\rho\sigma}\DAlembert\phi\\
& + 4R_{\nu\sigma\lambda\rho}\tensor{\phi}{^\lambda_\mu}\phi^{\rho\sigma} \big] \big\} + \frac{1}{2} G_{5\phi\phi X}\phi^\mu \big\{ 2\phi^\nu \left( \phi_{\lambda\nu}\tensor{\phi}{^\lambda_\mu}-\phi_{\mu\nu}\DAlembert\phi \right)+ \phi_\mu \left[ (\DAlembert\phi)^2 - (\phi_{\lambda\sigma})^2 \right] \big\} \\
& + \frac{1}{6} G_{5\phi XX} \phi^\mu \big\{ \phi_\mu \big[ 2(\phi^{\lambda\nu})^3+\DAlembert\phi \left( (\DAlembert\phi)^2-3(\phi_{\lambda\sigma})^2 \right) \big]+ 6\phi^\nu \big[ 2\tensor{\phi}{^\lambda_\mu}\big( \phi_{\lambda\nu}\DAlembert\phi-\phi_{\lambda\sigma}\tensor{\phi}{^\sigma_\nu} \big) \\
& - \phi_{\mu\nu} \left( (\DAlembert\phi)^2 - (\phi_{\lambda\sigma})^2  \right) \big] \big\}-\frac{1}{6} G_{5XXX}\phi^\mu \phi^\nu \big\{ \phi_{\mu\nu} \big[ 2 (\phi^{\lambda\sigma})^3+\DAlembert\phi \left( (\DAlembert\phi)^2-3(\phi_{\lambda\sigma})^2 \right) \big]\\
& +3\tensor{\phi}{^\lambda_\mu} \big[ 2\tensor{\phi}{^\sigma_\nu} \big(\phi_{\lambda\sigma}\DAlembert\phi  -\phi_{\rho\sigma} \tensor{\phi}{_\lambda^\rho} \big) -\phi_{\lambda\nu}\left( (\DAlembert\phi)^2-(\phi_{\rho\sigma})^2 \right) \big] \big\}=0.\\
\end{split}
\end{equation}

\end{widetext}

\section{Disformal invariance of the Horndeski lagrangian}\label{App:Disftransf}
The Horndeski Lagrangian~\eqref{Horndeski} is formally invariant under the transformation~\eqref{eq:disformal}~\cite{Bettoni:2013diz}. We derived independently these transformations and we found a mismatch with the results in~\cite{Bettoni:2013diz} which cannot be explained with differences in notation.
Formal invariance means that the Lagrangian maintains the same structure, upon redefinition of the free functions $G_i(\phi,X)$. For completeness, we report these transformations. Written with respect to the metric $\bar{g}_{\mu\nu}$, the Lagrangian reads
\begin{equation}
\label{HorndeskiBar}
\bar{S}=\frac{1}{2\kappa}\sum_{i=2}^{5}\int \dd^4x\,\sqrt{-\bar{g}}\bar{\mathcal{L}}_i,
\end{equation}
where we have defined
\begin{align}
\label{L2bar}
\bar{\mathcal{L}}_2 = & \, \bar{G}_2(\phi,\bar{X}),\\
\label{L3bar}
\bar{\mathcal{L}}_3 = & \, -\bar{G}_3(\phi,\bar{X})\bar{\DAlembert}\phi,\\
\label{L4bar}
\bar{\mathcal{L}}_4 = & \, \bar{G}_4(\phi,\bar{X})\bar{R} +\bar{G}_{4\bar{X}}[(\bar{\DAlembert}\phi)^2-(\bar{\nabla}_\mu\bar{\nabla}_\nu\phi)^2],\\
\notag
\bar{\mathcal{L}}_5 = & \, \bar{G}_5(\phi,\bar{X})\bar{G}_{\mu\nu}\bar{\nabla}^\mu\bar{\nabla}^\nu\phi \\
\label{L5bar}
- & \frac{\bar{G}_{5\bar{X}}}{6}\left[\left(\bar{\DAlembert}\phi\right)^3 -3\DAlembert\phi(\bar{\nabla}_\mu\bar{\nabla}_\nu\phi)^2+2(\bar{\nabla}_\mu\bar{\nabla}_\nu\phi)^3\right],
\end{align}
where the barred quantities are evaluated with the metric $\bar{g}_{\mu\nu}$. We can now define a new metric $g_{\mu\nu}$ which is related to $\bar{g}_{\mu\nu}$ through a disformal transformation
\begin{equation}\label{eq:disformal2}
\bar{g}_{\mu\nu} \equiv C(\phi)\left[g_{\mu\nu}+D(\phi)\nabla_\mu\phi\nabla_\nu\phi\right].
\end{equation}
As anticipated, under this transformation, Lagrangian~\eqref{HorndeskiBar} becomes Lagrangian~\eqref{Horndeski}, defined as in Eqs.~\eqref{L2}--\eqref{L5}. We can map the functions $G_i(\phi,X)$ in term of the barred functions $\bar{G}_i(\phi,\bar{X})$,
\begin{widetext}
\begin{subequations}
\begin{multline}\label{eq:G2transf}
G_2(\phi,X) = C^2\sqrt{1-2DX}\bar{G}_2(\phi,\bar{X}) + \frac{2X\bar{G}_3(\phi,\bar{X})}{\sqrt{1-2DX}}\left( C'+\frac{CD'X}{1-2DX}\right)+2XI_{3\phi} \\
+\frac{3X\bar{G}_4(\phi,\bar{X})}{\sqrt{1-2DX}}\left(-\frac{C'^2}{C}+2C''+\frac{2XC'D'}{1-2DX} \right)-4X\left[\bar{G}_4(\phi,\bar{X})\left(\frac{1+2D^2X^2}{\sqrt{1-2DX}}C'-CD'X\sqrt{1-2DX} \right) \right]_\phi  \\
+ \frac{12X^3C'D'\bar{G}_{4\bar{X}}(\phi,\bar{X})}{C(1-2DX)^{5/2}} +2XI_{4\phi} +\frac{3X^2C'\bar{G}_5(\phi,\bar{X})}{C^2(1-2DX)^{3/2}}\left(-\frac{2C'^2}{C}+2C''+\frac{3XC'D'}{1-2DX} \right) \\
+ \frac{2X^3C'^2\bar{G}_{5\bar{X}}(\phi,\bar{X})}{C^3(1-2DX)^{5/2}}\left(-\frac{C'}{C} +\frac{3XD'}{1-2DX}\right)-2X\left[\frac{X\bar{G}_5(\phi,\bar{X})}{\sqrt{1-2DX}}\left( \frac{(1+DX)C'^2}{(1-2DX)C^2}+\frac{C'D'X}{C}-\frac{2D'^2X^2}{1-2DX}\right)\right]_\phi  + 2XI_{5\phi},
\end{multline}

\begin{multline}\label{eq:G3transf}
	G_3(\phi,X) = \frac{C\bar{G}_3(\phi,\bar{X})}{\sqrt{1-2DX}} + I_3 -\frac{\bar{G}_4(\phi,\bar{X})}{\sqrt{1-2DX}}\left[ 4CD'X(1-2DX)-C'(5-4DX+4D^2X^2) \right]\\
+ \frac{2X\bar{G}_{4\bar{X}}}{(1-2DX)^{3/2}}\left[(1+2DX)\frac{C'}{C}+2D'X\right] +\frac{4CDX\bar{G}_{4\phi}(\phi,\bar{X})}{\sqrt{1-2DX}} + I_4 \\
-\frac{X\bar{G}_5}{\sqrt{1-2DX}}\left[ -\frac{C'^2}{2C^2}+\frac{XC'D'}{C}-\frac{4X^2D'^2(2-DX)}{(1-2DX)^2}-\frac{2XD''}{1-2DX} \right]\\
-\frac{X^2\bar{G}_{5\bar{X}}}{(1-2DX)^{5/2}}\left( -\frac{C'^2}{C^2}+\frac{2XC'D'}{C}-\frac{4X^2D'^2}{1-2DX} \right) -\frac{2X\bar{G}_{5\phi}}{(1-2DX)^{3/2}}\left( \frac{C'}{C}-XD' \right)
+I_5 +2XK_{5\phi\phi},
\end{multline}

\begin{equation}\label{eq:G4transf}
G_4(\phi,X) = C\sqrt{1-2DX}\bar{G}_4(\phi,\bar{X})+\frac{D'X^2\bar{G}_5(\phi,\bar{X})}{(1-2DX)^{3/2}} +XK_{5\phi},
\end{equation}

\begin{equation}\label{eq:G5transf}
G_5(\phi,X)=\frac{\bar{G}_5(\phi,\bar{X})}{\sqrt{1-2DX}}+K_5,
\end{equation}
\end{subequations}
where a prime or a subscript $\phi$ denotes a derivative with respect to $\phi$, a subscript $\bar{X}$ denotes a derivative with respect to $\bar{X}$, defined as
\begin{equation}
\bar{X}=-\frac{1}{2}\bar{g}^{\mu\nu}\partial_\mu\phi\partial_\nu\phi = \frac{X}{C(1-2DX)},
\end{equation}
and
\begin{gather}\label{eq:IntegralDisf}
I_3 = -CD\int\dd X\frac{\bar{G}_3(\phi,\bar{X})}{(1-2DX)^{3/2}}, \qquad I_4 = -\int\dd X \left[3\bar{G}_4(\phi,\bar{X})\sqrt{1-2DX}(CD)' +2\bar{G}_{4\phi}(\phi,\bar{X})\frac{CD}{\sqrt{1-2DX}}\right], \\
\begin{multlined}
I_5=-\int\dd X\bigg\{\frac{\bar{G}_5(\phi,\bar{X})}{(1-2DX)^{3/2}}\left[\frac{ (1-DX)C'^2}{2C^2}-\frac{(2-3DX)C'D'X}{C}+3D'^2X^2-D''X \right] \\
+ \frac{C'-CD'X}{C(1-2DX)^{3/2}}\bar{G}_{5\bar{X}}(\phi,\bar{X})-K_{5\phi\phi} \bigg\},
\end{multlined} \\
K_{5}=-D\int\dd X \frac{\bar{G}_5(\phi,\bar{X})}{(1-2DX)^{3/2}}.
\end{gather}
Our results of Eqs.~\eqref{eq:G2transf} and~\eqref{eq:G3transf} do not coincide with those of Eqs.~(C7) and~(C8) of Appendix C of~\cite{Bettoni:2013diz}.
\end{widetext}

\bibliographystyle{apsrev4-1}
\bibliography{bibnote}

%%%%
\end{document}